\documentstyle[12pt]{article}
\begin{document}
\tolerance=5000
\def\nn{\nonumber \\}
\def\be{\begin{equation}}
\def\ee{\end{equation}}
\def\bea{\begin{eqnarray}}
\def\eea{\end{eqnarray}}
\def\tr{{\rm tr}}
\def\cL{{\cal L}}
\def\e{{\rm e}}
\def\eg{{\it e.g.\ }}
\def\ie{{\it i.e.},\ }

\begin{flushright}
\begin{minipage}{3cm}
NDA-FP-23 \\
%hep-th/960xxxx
January (1996)
\end{minipage} 
\end{flushright}
\vfill
\begin{center}
{\large\bf Group Theoretical 
Structure of $N=1$ and $N=2$ Two-Form 
Supegravity}

\vfill

{\sc 
Shin'ichi NOJIRI}

\vfill

{\it Department of Mathematics and 
Physics}

{\it National Defence Academy}

{\it Hashirimizu, Yokosuka 139, JAPAN}

\vfill

{\bf ABSTRACT}
\end{center}

We clarifies the group theoretical 
structure of $N=1$ and $N=2$ two-form 
supergravity, which
is classically equivalent to the 
Einstein supergravity.
$N=1$ and $N=2$ two-form supergravity 
theories can be formulated as gauge theories.
By introducing two Grassmann variables $\theta^A$ 
($A=1,2$), we construct the explicit representations of 
the generators $Q^i$ of the gauge group, 
which makes to express any product of the 
generators as a linear combination 
of the generators $Q^iQ^j=\sum_k f^{ij}_k Q^k$.
By using the expression and the tensor product 
representation, we explain how to construct 
finite-dimensional representations of the gauge groups.
Based on these representations, we construct 
the Lagrangeans of $N=1$ and $N=2$ two-form  
supergravity theories.

\newpage

\section{Introduction}

The Einstein gravity theory 
might be an effective theory of 
a more fundamental theory \eg superstring theory
since the action of 
the Einstein gravity is not 
renormalizable.
%Recently, the non-perturbative structures 
%of the supersymmetric 
%gauge theory were clarified based 
%on the holomorphy and duality.
%The duality structure in the 
%effective supergravity theory
%of superstring theory 
%is also clarified and the clue for clarifying  
%the nonperturbative structure of 
%superstring theory is obtained.
Two-form gravity theory is known to be classically 
equivalent to the Einstein gravity theory and is obtained 
from a topological field theory, which is called BF 
theory  \cite{sch}, 
by imposing constraint conditions \cite{pl}.
The BF theory has a large local symmety called 
the Kalb-Ramond symmetry \cite{kara}.
Since the Kalb-Ramond symmetry is very stringy symmetry, 
the fundamental gravity theory is expected to be a kind of 
string theory \cite{kkns}.
%Since the two-form gravity theory is a chiral theory, 
%we could apply the non-perturbative analysis based on
%holomorphy or something.
The extension of the two-form gravity theory to the 
supergravity theory was considered in Ref.\cite{jacob}.
Furthermore the supergravity theory which has 
a cosmological term or $N=2$ supersymmetry was proposed in 
Ref.\cite{ks} and the group theoretical structure of 
these supergravity theory was discussed in Ref.\cite{ezawa}.
$N=1$ and $N=2$ two-form supergravity 
theories can be formulated as gauge theories.
In this paper, we call the gauge algebras as 
$N=1$ and $N=2$ topological superalgebras (TSA), 
which are the subalgebras of $N=1$ and $N=2$ 
Neveu-Schwarz algebaras whose generators are 
$(L_0, L_{\pm 1}, G_{\pm{1 \over 2}})$ and 
$(L_0, L_{\pm 1}, G^\pm_{\pm{1 \over 2}}, J_0)$, 
respectively.
$N=1$ topological superalgebra is nothing but 
$osp(1,2)$ algebra.

In this paper, 
by introducing two Grassmann variables $\theta^A$ 
($A=1,2$), we construct the explicit representations of 
the generators $Q^i$, which makes to express any product of the 
generators as a linear combination 
of the generators $Q^iQ^j=\sum_k f^{ij}_k Q^k$.
By using the expression and the direct product 
representation, we explain how to construct 
finite-dimensional representations of the gauge groups.
The representation theory makes it possible to construct 
a general action of two-form $N=1$ and $N=2$ 
supergravity theories, which is expected to give a clue for the 
non-perturbative analysis of the supergravity.
The non-perturbative analysis is also partially given 
in this paper.

This paper is organized as follows: In section 2, we give 
the representation theories of $N=1$ and $N=2$ topological 
superalgebra. By using the representation, we construct 
the Lagrangeans of $N=1$ and $N=2$ two-form supergravities in 
section 3. In section 4, we investigate the symmetry of the 
system and consider the non-perturbative effect. The last section 
is devoted to summary.

\section{The Representaions of $N=1$ and $N=2$ 
Topological Superalgebras}

By using two Grassmann (anti-commuting) 
variables $\theta^A$ 
($A=1,2$), we define the following generators,
\bea
\label{def1}
&& G_A={\partial \over \partial \theta^A},\ \ 
T^a=\theta^A T^{a\ B}_{\ A}
{\partial \over \partial \theta^B} \nn
&& I=\theta^A{\partial \over \partial \theta^A}, \ \ 
H^A=\theta^A\theta^B{\partial \over \partial \theta^B}
\eea
Here
\be
\label{T}
T^{a\ B}_{\ A}={1 \over 2}\sigma^{a\ B}_{\ A}
\ee
and $\sigma^a$'s ($a=1,2,3$) are Pauli matrices.
These generators make the following algebra, which 
we call topological superconformal algebra (TSCA)
in this paper:
\bea
\label{tsca}
&& \{G_A, H^B\}={1 \over 2}\delta_A^{\ B} I
-2T^{a\ B}_{\ A}T^a \nn
&& [G_A,T^a]=T^{a\ B}_{\ A}G_B, \ \ 
[T^a, H^A]=T^{a\ A}_{\ B}H^B \nn
&& [T^a,T^b]=i\epsilon^{abc}T^c, \ \ 
\{ G_A, G_B\}=\{H^A, H^B\}=0
\eea
This algebra contains  a closed subalgebra, 
which is found by defining an operator $\hat G_A$:
\be
\label{def2}
\hat G_A\equiv G_A+\alpha \epsilon_{AB} H^B
\ee
Here $\alpha$ is a parameter which can be absorbed 
into the redefinition of the operators but we keep  
$\alpha$ as a free parameter for later convenience.
Then $\hat G_A$ and $T^a$ make a closed algebra:
\bea
\label{tsa}
&&\{\hat G_A, \hat G_B\}=-4\alpha T^a_{AB}T^a \nn
&&[\hat G_A,T^a]=T^{a\ B}_{\ A} \hat G_B, \ \ 
[T^a,T^b]=i\epsilon^{abc}T^c
\eea
Here $T^a_{AB}$ is defined by
\be
\label{T2}
T^a_{AB}\equiv \epsilon_{BC}T^{a\ C}_{\ A}
\ee
and 
\bea
\label{eps}
&& \epsilon^{AB}=-\epsilon^{BA} \nn
&& \epsilon_{AB}=-\epsilon_{BA} \nn
&& \epsilon^{12}=\epsilon_{21}=1 \ .
\eea
The algebra (\ref{tsa}) is nothing but 
$osp(1,2)$ algebra which 
is the subalgebra of the Neveu-Schwarz 
algebra whose generators are $L_0$, $L_{\pm 1}$ and 
$G_{\pm {1 \over 2}}$.
In this paper, we call this algebra (\ref{tsa}) as 
$N=1$ topological superalgebra (TSA).
As we will see later,
$\hat G_A$ generates left-handed supersymmetry.

By defining the following operators,
\bea
\label{def3}
J&=&-{8 \over 3}\alpha T^aT^a
+\epsilon^{AB}\hat G_A\hat G_B \nn
G^1_A&=&\hat G_A \nn
G^2_A&=&{4 \over 3}iT^{a\ B}_{\ A} 
(T^a\hat G_B+\hat G_B T^a) \ ,
\eea
we can also construct an algebra, which we call 
$N=2$ topological superalgebra ($k,l=1,2$)
\bea
\label{tsa2}
&&\{G^k_A, \hat G^l_B\}=-4\delta^{kl}
\alpha T^a_{AB}T^a +i\epsilon^{kl}\epsilon_{AB}J\nn
&& [G^k_A,T^a]=T^{a\ B}_{\ A}G^k_B, \ \ 
[T^a,T^b]=i\epsilon^{abc}T^c \nn
&& [J,G^i_A]=i\alpha\epsilon^{ij}G^j_A,\ \ 
[T^a, J]=0
\eea
This algebra is the subalgebra of $N=2$ Neveu-Schwarz 
algebra whose generators are $L_0$, $L_{\pm 1}$, 
$G^{(\pm)}_{\pm {1 \over 2}}$ and $J_0$.

Since all the operators are explicitly given in terms 
of $\theta^A$ and ${\partial \over \partial \theta^A}$, 
we can find that the product of operators is given
by a linear combination of the operators:
\bea
\label{pro}
&&G^kG^l=\delta^{kl}
\left\{-2\alpha T^a_{AB}T^a
-\epsilon_{AB}\left({3 \over 2}J+2\alpha P\right)\right\} \nn
&& \hskip 1cm 
+i\epsilon^{kl}\left(-2\alpha T^a_{AB}T^a 
+{1 \over 2}\epsilon_{AB}J\right) \nn
&&T^aT^b={i \over 2}\epsilon^{abc}T^c+\delta^{ab}
\left({1 \over 4\alpha}J+{1 \over 2}P\right) \nn
&&J^2=3\alpha J+2\alpha^2P \nn
&&G^k_AT^a=T^aG^k_A+T^{a\ B}_AG^k_B
={1 \over 2}T^{a\ B}_AG^k_B-{i \over 2}\epsilon^{kl}
T^{a\ B}_AG^l_B \nn
&&JG^k_A=G^k_A J+i\alpha\epsilon^{kl}G^l_A
=-{1 \over 2}\alpha G^k_A+
{3 \over 2}i\alpha\epsilon^{kl}G^l_A \nn
&&JT^a=T^aJ=-\alpha T^a
\eea
Here $P$ is a projection operator
\be
\label{proj}
P^2=P
\ee
defined by
\be
\label{P}
P\equiv -{1 \over 2\alpha}\epsilon^{AB}
\hat G_A\hat G_B+2 T^a T^a
\ee
and $P$ acts as unity on the operators
\bea
\label{P2}
&& P G^k_A=G^k_AP=G^k_A \nn
&& P T^a=T^a P=T^a \nn
&& P J=J P=J
\eea
Therefore the invariant trace of the product 
of the operators can be defined by 
the coefficients of $P$ in Equation (\ref{pro}):
\bea
\label{tr}
&& \tr G^kG^l=-2\alpha\delta^{kl}\epsilon_{AB}, \ \ 
\tr T^aT^b={1 \over 2}\delta^{ab}, \ \ 
\tr J^2=2\alpha^2 \nn 
&& \tr G^k_AT^a=\tr T^aG^k_A=\tr JG^k_A= \tr G^k_A
\tr JT^a=\tr T^aJ=0
\eea
We express the product law (\ref{pro}) of the operators 
$Q^i$ ($Q^i=T^a$, $G^k_A$, $J$ and $P$) by
\be
\label{ex}
Q^iQ^j=\sum_k f^{ij}_k Q^k\ .
\ee
Especially the expression of the invariant trace (\ref{tr}) 
is given by
\be
\label{tr3}
\tr Q^iQ^j=g^{ij}\equiv f^{ij}_P
\ee

The representation of $N=1$ superalgebra is given by a 
doublet of the representations 
$\left(p,p+{1 \over 2}\right)$ 
in $SU(2)$ ($p$ is an integer or half-integer), 
which is generated by $T^a$ and that of $N=2$ 
is given by a quartet 
$\left(p,p+{1 \over 2},p+{1 \over 2},p+1\right)$.

$\left({1 \over 2}, 1\right)$ representation 
of $N=1$ superalgebra 
is giben by $(\hat G_A, T^a)$ and $\left(0,
{1 \over 2}\right)$ is given by $(J,G^2_A)$.
$(J,G^k_A,T^a)$ makes the $\left(0,{1 \over 2}, 
{1 \over 2}, 1\right)$ representation of $N=2$ 
superalgebra.

$\left(1,{3 \over 2}\right)$ and 
$\left({3 \over 2},2\right)$ representations 
of $N=1$ superalgebra are 
given by a tensor product, where $\hat G_A$ and 
$T^a$ are replaced 
by $\hat G_A\otimes P + P\otimes \hat G_A$ and 
$T^a \otimes P + P \otimes T^a$:
\begin{itemize} 
\item $\left(1,{3 \over 2}\right)$ reprezentation 
$(K_{AB},N^2_{ABC})$:
\bea
\label{KN2}
K_{AB}&=&-T^a_{AB}(J\otimes T^a+T^a\otimes J)
-i{1 \over 2}\epsilon_{kl}G^k_{(A}\otimes G^l_{B)} \nn
N^2_{ABC}&=&T^a_{(AB}(G^2_{C)}\otimes T^a
+T^a \otimes G^2_{C)})
\eea
Here $(AB\cdots X)$ means a symmetrization with respect 
to the indeces $AB\cdots X$.
\item $\left({3 \over 2},2\right)$ representation 
$(N^1_{ABC}, M_{ABCD})$:
\bea
\label{N1M}
N^1_{ABC}&=&T^a_{(AB}(G^1_{C)}\otimes T^a
+T^a \otimes G^1_{C)}) \nn
M_{ABCD}&=&T^a_{(AB}T^b_{CD)}T^a\otimes T^b
\eea
\end{itemize}
$(K_{AB},N^k_{ABC}, M_{ABCD})$ makes 
$\left(1,{3 \over 2}, {3 \over 2},2\right)$
representation of $N=2$ superalgebra.
The commutater of $G^k_A$ with 
$(K_{AB},N^k_{ABC}, M_{ABCD})$ are given by\footnote{
$G_A^k$ in Equation (\ref{KNNM}) 
is understood to be $G_A^k\otimes P+P\otimes G_A^k$.}
\bea
\label{KNNM}
&& [G^k_E, M_{ABCD}]=-{1 \over 2}\epsilon_{E(A}
N^k_{BCD)} \nn
&& \{G^k_D, N^l_{ABC}\}=-8\alpha \delta^{kl}M_{ABCD}
-i\epsilon^{kl}\epsilon_{D(A}K_{BC)} \nn
&& [G^k_C, K_{AB}]=-3i\epsilon^{kl}N^l_{ABC} \ .
\eea
The coefficient of $P\otimes P$ in the product of 
$K_{AB}$, $N^k_{ABC}$ and $M_{ABCD}$ gives the
invariant trace
\bea
\label{tr2}
\tr M_{ABCD}M_{A'B'C'D'}&=&
{1 \over 2}\epsilon_{A(A'}\epsilon_{\hat B B'}
\epsilon_{\hat C C'}\epsilon_{\hat D D')} \nn
\tr N^k_{ABC}N^l_{A'B'C'}&=&\alpha 
\epsilon_{A(A'}\epsilon_{\hat B B'}\epsilon_{\hat C C')} \nn
\tr K_{AB}K_{A'B'}&=& \alpha^2
\epsilon_{A(A'}\epsilon_{\hat B B')} 
\eea
Here $(AB\cdots \hat F \cdots Z)$ means the symmetrization
with respect to the indeces $AB\cdots Z$ {\it except} $F$.

\section{The Lagrangeans of $N=1$ and $N=2$ Two-Form 
Supergravities}

In order to construct the Lagrangean of 
$N=1$ two-form supergravity theory, 
we introduce the gauge field $A_\mu$ which is 
$\left({1 \over 2}, 1\right)$ representation
\be
\label{gf1}
A_\mu=\psi_\mu^A \hat G_A+\omega^a_\mu T^a
\ee
and define the field strength as follows
\bea
\label{fs1}
R_{\mu\nu}&=&[\partial_\mu+A_\mu,\partial_\nu+A_\nu] \nn
&=&\{\partial_\mu\psi_\nu^A-\partial_\nu\psi_\mu^A
+T^{a\ A}_B(\psi_\mu^B\omega_\nu^a-\psi_\nu^B\omega_\mu^a)
\}\hat G_A \nn
&& +\{\partial_\mu\omega_\nu^a-\partial_\nu\omega_\mu^a
+i\epsilon^{abc}\omega_\mu^b\omega_\nu^c
+4\alpha T^a_{AB}\psi_\mu^A\psi_\nu^B
\}T^a
\eea
The left-handed supersymmetry transformation law of 
the gauge fields is given by
\bea
\label{gst}
\delta_G A_\mu&=&[\epsilon^A\hat G_A, \partial_\mu
+A_\mu] \nn
&=&(-\partial_\mu\epsilon^A +T^{a\ A}_B\epsilon^B
\omega_\mu^a)\hat G_A 
+4\alpha T^a_{AB}\epsilon^A\psi_\mu^BT^a
\eea

We also introduce the two-form field $X_{\mu\nu}$ 
which is $\left({1 \over 2}, 1\right)$ 
representation:
\be
\label{two1}
X_{\mu\nu}=\chi_{\mu\nu}^A \hat G_A+\Sigma^a_{\mu\nu} T^a
\ee
Then the Lagrangean ${\cal L}_{\rm BF}$ 
of the so-called BF theory with 
$N=1$ local supersymmetry is given by
\bea
\label{KR1}
{\cal L}_{\rm BF}&=&\epsilon^{\mu\nu\rho\sigma}\left\{
{1 \over g}\tr R_{\mu\nu}X_{\rho\sigma}
+\Lambda \tr X_{\mu\nu}X_{\rho\sigma} \right\} \\
\tr R_{\mu\nu}X_{\rho\sigma} &
=&2\alpha\epsilon_{AB}
\{\partial_\mu\psi_\nu^A-\partial_\nu\psi_\mu^A
+T^{a\ A}_B(\psi_\mu^B\omega_\nu^a-\psi_\nu^B\omega_\mu^a)
\}\chi_{\rho\sigma}^B \nn
&& +{1 \over 2}
\{\partial_\mu\omega_\nu^a-\partial_\nu\omega_\mu^a
+i\epsilon^{abc}\omega_\mu^b\omega_\nu^c
+4\alpha T^a_{AB}\psi_\mu^A\psi_\nu^B
\}\Sigma^a_{\rho\sigma} \\
\label{cosmo1}
\tr X_{\mu\nu}X_{\rho\sigma}&=&
2\alpha\epsilon_{AB}\chi_{\mu\nu}^A\chi_{\rho\sigma}^B
+{1 \over 2}\Sigma^a_{\mu\nu}\Sigma^a_{\rho\sigma}
\eea
Here $g$ is a gauge coupling constant and 
$\Lambda$ is a cosmological constant.
In order to obtain $N=1$ two-form 
supergravity theory, we need to introduce the 
multiplier field $\Phi$ which is 
$\left({3 \over 2}, 2\right)$ representation:
\be
\label{mul1}
\Phi=\kappa^{ABC}N_{ABC}+\phi^{ABCD}M_{ABCD}
\ee
The Lagrangean ${\cal L}$ of $N=1$ two-form supergravity is 
given by adding the constraint term to 
the Lagrangean ${\cal L}_{\rm BF}$:
\bea
\label{cons1}
{\cal L}&=&{\cal L}_{\rm BF}
+\epsilon^{\mu\nu\rho\sigma}
\tr \Phi(X_{\mu\nu}\otimes X_{\rho\sigma}) \nn
\tr \Phi(X_{\mu\nu}\otimes X_{\rho\sigma})&=&
-\alpha T^a_{AB}\epsilon_{CD}\kappa^{ABC}
(\chi_{\mu\nu}^D\Sigma^a_{\rho\sigma}
+\Sigma^a_{\mu\nu}\chi^D_{\rho\sigma}) \nn
&& +T^a_{AB}T^b_{CD}\phi^{ABCD}
\Sigma_{\mu\nu}^a\Sigma_{\rho\sigma}^b
\eea

The Lagrangean of $N=2$ theory is also given by 
introducing gauge field which is  
$\left(0, {1 \over 2}, {1 \over 2}, 1\right)$ 
representation
\be
\label{gf2}
A_\mu=B_\mu J+\psi^{kA}_\mu G^k_A+\omega^a_\mu T^a
\ee
and defining the field strength
\bea
\label{fs2}
R_{\mu\nu}&=&[\partial_\mu+A_\mu,\partial_\nu+A_\nu] \nn
&=&\{\partial_\mu B_\nu-\partial_\nu B_\mu
-i\epsilon_{AB}\epsilon^{kl}
\psi^{k\ A}_\mu
\psi^{l\ B}_\nu\}J \nn
&&+
\{\partial_\mu\psi^{k\ A}_\nu-\partial_\nu\psi_\mu^{k\ A}
+T^{a\ A}_B(\psi_\mu^{k\ B}
\omega_\nu^a-\psi_\nu^{k\ B}\omega_\mu^a) \nn
&&-i\epsilon^{kl}(B_\mu\psi^{l\ A}_\nu-B_\nu\psi^{l\ A}_\nu
)\}\hat G_A \nn
&& +\{\partial_\mu\omega_\nu^a-\partial_\nu\omega_\mu^a
+i\epsilon^{abc}\omega_\mu^b\omega_\nu^c
+4\alpha T^a_{AB}\psi_\mu^{k\ A}\psi_\nu^{k\ B}
\}T^a
\eea
The gauge transformation law of the gauge field has 
the following form:
\bea
\label{gt}
\delta A_\mu&=&[aJ+\epsilon^{kA} G^k_A
+\delta^aT^a, \partial_\mu+A_\mu] \nn
&=&(-\partial_\mu a 
-i\epsilon_{AB}\epsilon^{kl}\epsilon^{kA}\psi_\mu^{lB})J \nn
&&+\left\{-\partial_\mu\epsilon^{kA} +T^{a\ A}_B(\epsilon^{kB}
\omega_\mu^a+i\delta^a\psi_\mu^{kB})
-i\alpha\epsilon^{kl}(a\psi_\mu^{lA}
-\epsilon^{lA}B_\mu)\right\}G^k_A \nn
&&+(-\partial_\mu\delta^a
+i\epsilon^{abc}\delta^b\omega_\mu^c
+4\alpha T^a_{AB}\epsilon^{kA}\psi^{kB}_\mu)T^a
\eea
The two-form field in $N=2$ theory is 
$\left(0, {1 \over 2}, {1 \over 2}, 1\right)$
representation
\be
\label{two2}
X_{\mu\nu}=\Pi_{\mu\nu}J
+\chi_{\mu\nu}^{kA}G^k_A +\Sigma^a_{\mu\nu} T^a
\ee
Then the Lagrangean of $N=2$ BF theory is given by
\bea
\label{KR2}
{\cal L}_{\rm BF}&=&\epsilon^{\mu\nu\rho\sigma}\left\{
{1 \over g}\tr R_{\mu\nu}X_{\rho\sigma}
+\Lambda \tr X_{\mu\nu}X_{\rho\sigma} \right\} \\
\tr R_{\mu\nu}X_{\rho\sigma}&=&2\alpha^2
\{\partial_\mu B_\nu-\partial_\nu B_\mu
-i\epsilon_{AB}\epsilon^{kl}
\psi^{k\ A}_\mu
\psi^{l\ B}_\nu\}\Pi_{\rho\sigma} \nn
&&+2\alpha
\{\partial_\mu\psi^{k\ A}_\nu-\partial_\nu\psi_\mu^{k\ A}
+T^{a\ A}_B(\psi_\mu^{k\ B}
\omega_\nu^a-\psi_\nu^{k\ B}\omega_\mu^a) \nn
&&
-i\epsilon^{kl}(B_\mu\psi^{l\ A}_\nu-B_\nu\psi^{l\ A}_\nu
)\}\chi^{kB}_{\rho\sigma} \nn
&& +{1 \over 2}
\{\partial_\mu\omega_\nu^a-\partial_\nu\omega_\mu^a
+i\epsilon^{abc}\omega_\mu^b\omega_\nu^c
+4\alpha T^a_{AB}\psi_\mu^{k\ A}\psi_\nu^{k\ B}
\}\Sigma^a_{\rho\sigma} \\
\label{cosmo2}
\tr X_{\mu\nu}X_{\rho\sigma}&=&
2\alpha^2\Pi_{\mu\nu}\Pi_{\rho\sigma}
+2\alpha\epsilon_{AB}\chi_{\mu\nu}^{kA}
\chi_{\rho\sigma}^{kB}
+{1 \over 2}\Sigma^a_{\mu\nu}\Sigma^a_{\rho\sigma}
\eea
The Lagrangean ${\cal L}$ of $N=2$ two-form supergravity theory is 
given by introducing the multiplier field which is  
$\left(1, {3 \over 2}, {3 \over 2}, 2\right)$ 
representation
\be
\label{mul2}
\Phi=\lambda^{AB}K_{AB}
+\kappa^{kABC}N^k_{ABC}+\phi^{ABCD}M_{ABCD}
\ee
and adding the term which gives the constraint on 
the two-form field
\bea
\label{cons2}
{\cal L}&=&{\cal L}_{\rm BF}
+\epsilon^{\mu\nu\rho\sigma}
\tr \Phi(X_{\mu\nu}\otimes X_{\rho\sigma}) \\
\tr \Phi(X_{\mu\nu}\otimes X_{\rho\sigma})&=&
\alpha^2\lambda^{AB}\{-T^a_{AB}(\Pi_{\mu\nu}
\Sigma^a_{\rho\sigma}+\Sigma^a_{\mu\nu}\Pi_{\rho\sigma})
+i\epsilon^{kl}\epsilon_{AC}\epsilon_{BD}
\chi^{kC}_{\mu\nu}\chi^{lD}_{\rho\sigma}\} \nn
&&-\alpha T^a_{AB}\epsilon_{CD}\kappa^{kABC}
(\chi^{kD}_{\mu\nu}\Sigma^a_{\rho\sigma}
+\Sigma^a_{\mu\nu}\chi^{kD}_{\rho\sigma}) \nn
&& +T^a_{AB}T^b_{CD}\phi^{ABCD}
\Sigma_{\mu\nu}^a\Sigma_{\rho\sigma}^b
\eea

\section{The Symmetry of the Lagrangeans}

We now consider the right-handed supersymmetry.
The Lagrangeans of the $N=1$ and $N=2$ have the 
following form
\be
\label{lag0}
\cL=\epsilon^{\mu\nu\rho\sigma}\left\{
{1 \over g}\tr R_{\mu\nu}X_{\rho\sigma} 
+\Lambda\tr X_{\mu\nu} X_{\rho\sigma}
+\tr \Phi(X_{\mu\nu} \otimes X_{\rho\sigma})
\right\}
\ee
On the other hand 
the Lagrangeans of the corresponding BF theory
have the following form
\be
\label{lagKR}
\cL_{\rm BF}=\epsilon^{\mu\nu\rho\sigma}\left\{
{1 \over g}\tr R_{\mu\nu}X_{\rho\sigma} 
+\Lambda\tr X_{\mu\nu} X_{\rho\sigma}
\right\}
\ee
The Lagrangean (\ref{lagKR}) has the large local 
symmetry which is called Kalb-Ramond symmetry.
The parameter of the transformation $C_\mu$ is 
$\left({1 \over 2}, 1\right)$
representation in $N=1$ theory and 
$\left(0, {1 \over 2}, {1 \over 2}, 1\right)$
representation in $N=2$ theory and the 
transformation law of the Kalb-Ramond symmetry is given 
by 
\bea
\label{KRtr1}
\delta_{{\rm KR}} A_\mu&=& -g\Lambda C_\mu \nn
\delta_{{\rm KR}} X_{\mu\nu}&=&{1 \over 2}(D_\mu C_\nu-
D_\nu C_\mu) 
\eea
Here the covariant derivative $D_\mu$ is defined by
\be
D_\mu \ \cdot\ =[\partial_\mu+A_\mu,\ \cdot \ ]
\ee
Now we consider the Kalb-Ramond like transformation 
for the Lagrangean (\ref{lag0}):
\bea
\label{KRtr2}
\delta_{{\rm KR}} A_\mu&=& -g\Lambda C_\mu 
-g \Phi\times C_\mu \nn
\delta_{{\rm KR}} X_{\mu\nu}&=&{1 \over 2}(D_\mu C_\nu-
D_\nu C_\mu) 
\eea
Here the product $R\times S$ of two operators 
$R=\sum_{ij} r_{ij}Q^i\otimes Q^j$, 
which is $\left({3 \over 2}, 2\right)$
representation in $N=1$ theory and 
$\left(1, {3 \over 2}, {3 \over 2}, 2\right)$
representation in $N=2$ theory, and $S=\sum_i s_i Q^i$, 
which is $\left({1 \over 2}, 1\right)$
representation in $N=1$ theory and 
$\left(0, {1 \over 2}, {1 \over 2}, 1\right)$
representation in $N=2$ theory, 
is defined by
\be
\label{pr}
R \times S \equiv \sum_{ijk} s_i r_{jk} g^{ik} G^j
\ee
Here $g^{ik}$ is defined in Equation (\ref{tr3}).
The product $R\times S$ is $\left({3 \over 2}, 2\right)$
representation in $N=1$ theory and 
$\left(1, {3 \over 2}, {3 \over 2}, 2\right)$
representation in $N=2$ theory.
Then the change of the Lagrangean (\ref{lag0}) is given by
\be
\label{change}
\delta_{{\rm KR}} \cL = -\epsilon^{\mu\nu\rho\sigma}
\tr D_\mu  \Phi C_\nu \Sigma_{\rho\sigma} 
+{\rm \ total\ derivative}
\ee
This tells that the Lagrangean (\ref{lag0}) is invariant
if the parameter $C_\mu$ satisfies the equation
\be
\label{KRcon}
0=\epsilon^{\mu\nu\rho\sigma}
B_\nu \otimes \Sigma_{\rho\sigma}|_{\left({3 \over 2},2\right)\ 
{\rm or}\ \left(1,{3 \over 2},{3 \over 2},2\right)\ {\rm part}}\ .
\ee
Equation (\ref{KRcon}) has non-trivial solutions and 
the fermionic part of the solution 
corresponds to right-handed supersymmetry \cite{ks}.
The commutator of the right-handed supersymmetry transformation 
and the left-handed one contains the general coordinate 
transformation.

When $\alpha\neq 0$, the parameter $\alpha$ 
can be absorbed into the redefinition 
of the operators or fields as follows:
\bea
\label{redef1}
&&\omega^a_\mu\rightarrow\omega^a_\mu,\ \ 
\psi^{kA}_\mu\rightarrow\alpha^{-{1 \over 2}}
\psi^{kA}_\mu, \ \ 
B_\mu\rightarrow\alpha^{-1}B_\mu \nn
&&\Sigma^a_{\mu\nu}\rightarrow\Sigma^a_{\mu\nu}, \ \ 
\chi^{kA}_{\mu\nu}\rightarrow
\alpha^{-{1 \over 2}}\chi^{kA}_{\mu\nu}, \ \ 
\Pi_{\mu\nu}\rightarrow\alpha^{-1}\Pi_{\mu\nu} \nn
&&\phi^{ABCD}\rightarrow\phi^{ABCD}, \ \ 
\kappa^{ABC}\rightarrow
\alpha^{-{1 \over 2}}\kappa^{ABC}, \ \ 
\lambda^{AB}\rightarrow\alpha^{-1}\lambda^{AB} \ .
\eea
Then $N=1$ Lagrangean has the following form
\bea
\label{lag3}
\cL&=&\epsilon^{\mu\nu\rho\sigma}\Bigl[
{1 \over g}\Bigl\{
2\epsilon_{AB}
\{\partial_\mu\psi_\nu^A-\partial_\nu\psi_\mu^A
+T^{a\ A}_B(\psi_\mu^B\omega_\nu^a-\psi_\nu^B\omega_\mu^a)
\}\chi_{\rho\sigma}^B \nn
&& +{1 \over 2}
(\partial_\mu\omega_\nu^a-\partial_\nu\omega_\mu^a
+i\epsilon^{abc}\omega_\mu^b\omega_\nu^c
+4T^a_{AB}\psi_\mu^A\psi_\nu^B)\Sigma^a_{\rho\sigma}\Bigr\} \nn
&&+\Lambda\left\{2
\epsilon_{AB}\chi_{\mu\nu}^A\chi_{\rho\sigma}^B
+{1 \over 2}
\Sigma^a_{\mu\nu}\Sigma^a_{\rho\sigma}\right\}\nn
&&- T^a_{AB}\epsilon_{CD}\kappa^{ABC}
(\chi^{D}_{\mu\nu}\Sigma^a_{\rho\sigma}
+\Sigma^a_{\mu\nu}\chi^{D}_{\rho\sigma}) \nn
&& +T^a_{AB}T^b_{CD}\phi^{ABCD}
\Sigma_{\mu\nu}^a\Sigma_{\rho\sigma}^b\Bigr]
\eea
and $N=2$ Lagrangean the following form
\bea
\label{lag4}
\cL&=&\epsilon^{\mu\nu\rho\sigma}\Bigl[{1 \over g}\Bigl\{
2\{\partial_\mu B_\nu-\partial_\nu B_\mu
-i\epsilon_{AB}\epsilon^{kl}
\psi^{k\ A}_\mu\psi^{l\ B}_\nu\}\Pi_{\rho\sigma} \nn
&&+2
\{\partial_\mu\psi^{k\ A}_\nu-\partial_\nu\psi_\mu^{k\ A}
+T^{a\ A}_B(\psi_\mu^{k\ B}
\omega_\nu^a-\psi_\nu^{k\ B}\omega_\mu^a) \nn
&& 
-i\epsilon^{kl}(B_\mu\psi^{l\ A}_\nu-B_\nu\psi^{l\ A}_\nu
)\}\chi^{kB}_{\rho\sigma} \nn
&& +{1 \over 2}
(\partial_\mu\omega_\nu^a-\partial_\nu\omega_\mu^a
+i\epsilon^{abc}\omega_\mu^b\omega_\nu^c
+4T^a_{AB}\psi_\mu^{k\ A}\psi_\nu^{k\ B}
)\Sigma^a_{\rho\sigma}\Bigr\} \nn
&&+\Lambda\Bigl\{2\Pi_{\mu\nu}\Pi_{\rho\sigma}
+2\epsilon_{AB}\chi_{\mu\nu}^{kA}
\chi_{\rho\sigma}^{kB}
+{1 \over 2}\Sigma^a_{\mu\nu}\Sigma^a_{\rho\sigma}
\Bigr\}\nn
&&+\lambda^{AB}\{-T^a_{AB}(\Pi_{\mu\nu}
\Sigma^a_{\rho\sigma}+\Sigma^a_{\mu\nu}\Pi_{\rho\sigma})
+i\epsilon^{kl}\epsilon_{AC}\epsilon_{BD}
\chi^{kC}_{\mu\nu}\chi^{lD}_{\rho\sigma}\} \nn
&&- T^a_{AB}\epsilon_{CD}\kappa^{kABC}
(\chi^{kD}_{\mu\nu}\Sigma^a_{\rho\sigma}
+\Sigma^a_{\mu\nu}\chi^{kD}_{\rho\sigma}) \nn
&& +T^a_{AB}T^b_{CD}\phi^{ABCD}
\Sigma_{\mu\nu}^a\Sigma_{\rho\sigma}^b\Bigr]
\eea
The Lagrangeans (\ref{lag3}) and (\ref{lag4}) are 
nothing but the Lagrangeans found in Ref.\cite{ks}. 
These Lagrangeans are invariant 
under the following $U(1)$ ``symmetry''
\bea
\label{u1}
&&\omega^a_\mu\rightarrow\omega^a_\mu,\ \ 
\psi^{kA}_\mu\rightarrow\psi^{kA}_\mu, \ \ 
B_\mu\rightarrow B_\mu \nn
&&\Sigma^a_{\mu\nu}\rightarrow\e^\varphi
\Sigma^a_{\mu\nu}, \ \ 
\chi^{kA}_{\mu\nu}\rightarrow
\e^\varphi\chi^{kA}_{\mu\nu}, \ \ 
\Pi_{\mu\nu}\rightarrow\e^\varphi\Pi_{\mu\nu}\nn
&&\phi^{ABCD}\rightarrow\e^{-2\varphi}\phi^{ABCD}, \ \ 
\kappa^{ABC}\rightarrow
\e^{-2\varphi}\kappa^{ABC}, \ \ 
\lambda^{AB}\rightarrow\e^{-2\varphi}\lambda^{AB}\nn
&&, g\rightarrow\e^\varphi g
 ,\ \
\Lambda\rightarrow\e^{-2\varphi}\Lambda
\eea

We can also consider $\alpha\rightarrow 0$ theory
by redefining the fields as follows
\bea
\label{redef2}
&&\omega^a_\mu\rightarrow\omega^a_\mu,\ \ 
\psi^{kA}_\mu\rightarrow\psi^{kA}_\mu, \ \ 
B_\mu\rightarrow B_\mu \nn
&&\Sigma^a_{\mu\nu}\rightarrow\Sigma^a_{\mu\nu}, \ \ 
\chi^{kA}_{\mu\nu}\rightarrow
\alpha^{-1}\chi^{kA}_{\mu\nu}, \ \ 
\Pi_{\mu\nu}\rightarrow\alpha^{-2}\Pi_{\mu\nu} \nn
&&\phi^{ABCD}\rightarrow\phi^{ABCD}, \ \ 
\kappa^{ABC}\rightarrow\kappa^{ABC}, \ \ 
\lambda^{AB}\rightarrow\lambda^{AB} \nn
&& g\rightarrow g,\ \ 
\Lambda\rightarrow
\Bigl\{\begin{array}{ll}\alpha\Lambda & (N=1) \\
\alpha^2\Lambda & (N=2) \end{array} 
\eea
then $N=1$ Lagrangean is rewritten by
\bea
\label{lag5}
\cL&=&\epsilon^{\mu\nu\rho\sigma}
\Bigl[{1 \over g}\Bigl\{
2\epsilon_{AB}
\{\partial_\mu\psi_\nu^A-\partial_\nu\psi_\mu^A
+T^{a\ A}_B(\psi_\mu^B\omega_\nu^a-\psi_\nu^B\omega_\mu^a)
\}\chi_{\rho\sigma}^B \nn
&& +{1 \over 2}
(\partial_\mu\omega_\nu^a-\partial_\nu\omega_\mu^a
+i\epsilon^{abc}\omega_\mu^b\omega_\nu^c
)\Sigma^a_{\rho\sigma}\Bigr\} \nn
&&+2\Lambda \epsilon_{AB}
\chi_{\mu\nu}^A\chi_{\rho\sigma}^B
\nn
&&- T^a_{AB}\epsilon_{CD}\kappa^{ABC}
(\chi^{D}_{\mu\nu}\Sigma^a_{\rho\sigma}
+\Sigma^a_{\mu\nu}\chi^{D}_{\rho\sigma}) \nn
&& +T^a_{AB}T^b_{CD}\phi^{ABCD}
\Sigma_{\mu\nu}^a\Sigma_{\rho\sigma}^b
\eea
The above Lagrangean with $\Lambda=0$ 
was found in Ref.\cite{jacob}.
On the other hand the $N=2$ Lagrangean has the following form: 
\bea
\label{lag6}
\cL&=&\epsilon^{\mu\nu\rho\sigma}\Bigl[{1 \over g}\Bigl\{
2\{\partial_\mu B_\nu-\partial_\nu B_\mu
-i\epsilon_{AB}\epsilon^{kl}
\psi^{k\ A}_\mu\psi^{l\ B}_\nu\}\Pi_{\rho\sigma} \nn
&&+2
\{\partial_\mu\psi^{k\ A}_\nu-\partial_\nu\psi_\mu^{k\ A}
+T^{a\ A}_B(\psi_\mu^{k\ B}
\omega_\nu^a-\psi_\nu^{k\ B}\omega_\mu^a)
\}\chi^{kB}_{\rho\sigma} \nn
&& +{1 \over 2}
(\partial_\mu\omega_\nu^a-\partial_\nu\omega_\mu^a
+i\epsilon^{abc}\omega_\mu^b\omega_\nu^c
)\Sigma^a_{\rho\sigma} \\
&&+2\Lambda\Pi_{\mu\nu}\Pi_{\rho\sigma}
\nn
&&+\lambda^{AB}\{-T^a_{AB}(\Pi_{\mu\nu}
\Sigma^a_{\rho\sigma}+\Sigma^a_{\mu\nu}\Pi_{\rho\sigma})
+i\epsilon^{kl}\epsilon_{AC}\epsilon_{BD}
\chi^{kC}_{\mu\nu}\chi^{lD}_{\rho\sigma}\} \nn
&&- T^a_{AB}\epsilon_{CD}\kappa^{kABC}
(\chi^{kD}_{\mu\nu}\Sigma^a_{\rho\sigma}
+\Sigma^a_{\mu\nu}\chi^{kD}_{\rho\sigma}) \nn
&& +T^a_{AB}T^b_{CD}\phi^{ABCD}
\Sigma_{\mu\nu}^a\Sigma_{\rho\sigma}^b
\eea
The Lagrangeans (\ref{lag5}) and (\ref{lag6}) have two kinds 
of $U(1)$ ``symmetries'', one of which is given by
\bea
\label{u12}
&&\omega^a_\mu\rightarrow\omega^a_\mu,\ \ 
\psi^{kA}_\mu\rightarrow\psi^{kA}_\mu, \ \ 
B_\mu\rightarrow B_\mu\nn
&&\Sigma^a_{\mu\nu}\rightarrow\e^\varphi
\Sigma^a_{\mu\nu}, \ \ 
\chi^{kA}_{\mu\nu}\rightarrow
\e^\varphi\chi^{kA}_{\mu\nu}, \ \ 
\Pi_{\mu\nu}\rightarrow\e^\varphi\Pi_{\mu\nu}\nn
&&\phi^{ABCD}\rightarrow\e^{-2\varphi}\phi^{ABCD}, \ \ 
\kappa^{ABC}\rightarrow
\e^{-2\varphi}\kappa^{ABC}, \ \ 
\lambda^{AB}\rightarrow\e^{-2\varphi}\lambda^{AB}\nn
&& g\rightarrow\e^\varphi g ,\ \
\Lambda\rightarrow\e^{-2\varphi}\Lambda
\eea
We call the another $U(1)$ ``symmetry'' as 
 $U(1)_R$ ``symmetry'' since the symmetry corresponding to 
the scale transformation of the Grassmann number $\theta^A$.
The $U(1)_R$ ``symmetry'' is given by
\bea
\label{u1R}
&&\omega^a_\mu\rightarrow\omega^a_\mu,\ \ 
\psi^{kA}_\mu\rightarrow\e^{\rho}\psi^{kA}_\mu, \ \ 
B_\mu\rightarrow \e^{2\rho}B_\mu\nn
&&\Sigma^a_{\mu\nu}\rightarrow \Sigma^a_{\mu\nu}, \ \ 
\chi^{kA}_{\mu\nu}\rightarrow\e^{-\rho}
\chi^{kA}_{\mu\nu}, \ \ 
\Pi_{\mu\nu}\rightarrow\e^{-2\rho}\Pi_{\mu\nu} \nn
&&\phi^{ABCD}\rightarrow\phi^{ABCD}, \ \ 
\kappa^{ABC}\rightarrow\e^{\rho}\kappa^{ABC}, \ \ 
\lambda^{AB}\rightarrow\e^{2\rho}\lambda^{AB}\nn
&& g\rightarrow g,\ \ 
\Lambda\rightarrow
\Bigl\{\begin{array}{ll}\e^{2\rho}\Lambda & (N=1) \\
\e^{4\rho}\Lambda & (N=2) \end{array}
\eea

If we assume the above $U(1)$ symmetries survive in 
the quantum theory, the form of the effective Lagrangean is  
restricted.
If we started from the theory which does not has a cosmological 
term ($\Lambda=0$), 
the gauge symmetry including the left-handed supersymmetry 
restricts the form of the terms appearing 
in the effective Lagrangean 
as $g^l\left({1 \over g} R\right)^m X^n$ after integrating 
the multiplier field $\Phi$ (Here we abbreviated the Lorentz 
indeces). The $U(1)$ symmetry and Lorentz symmetry give the
further restrictions:
\bea
\label{eff}
l-m+n&=&0 \\
m+n&=&2
\eea
\ie 
\be
\label{eff2}
l=2m-2
\ee
It would be natural to assume the theory has the good weak 
coupling limit ($g\rightarrow 0$), which gives $l\geq 0$.
We also assume $m\geq 0$ since $R$ contains the derivative.
Then there does not appear the cosmological term even 
in the quantum theory.
The term proportional to $R^m$ appears only perturbatively.
Since there does not appear the higher derivative terms 
perturbatively, the possible terms are 
$(l,m,n)=(0,1,1)$, $(2,2,0)$. 
Therefore if the term of $(l,m,n)=(2,2,0)$ do not appear at the 
order of $g^2$, only the term in the original Lagrangean \ie 
the term of $(l,m,n)=(0,1,1)$ can appear.
This might tell only that there is no quantum correction and 
the Einstein theory is the unique infrared theory.

\section{Summary}

In this paper, we have considered the group theoretical 
structure of $N=1$ and $N=2$ two-form supergravity 
theories based on 
$N=1$ and $N=2$ topological superalgebras (TSA), 
which are the subalgebras of $N=1$ and $N=2$ 
Neveu-Schwarz algebaras whose generators are 
$(L_0, L_{\pm 1}, G_{\pm{1 \over 2}})$ and 
$(L_0, L_{\pm 1}, G^\pm_{\pm{1 \over 2}}, J_0)$, 
respectively.
By introducing two Grassmann variables $\theta^A$ 
($A=1,2$), we have found the explicit representations of 
the generators $Q^i$ and we found that any product of the 
generators is given by a linear combination 
of the generators; $Q^iQ^j=\sum_k f^{ij}_k Q^k$.
By using the expression and the direct product 
representation, it has been explained how to construct 
finite-dimensional representation of the gauge groups.
It is expected that this gives a clue for the 
non-perturbative analysis of the supergravity.

\vskip 1cm

\noindent
{\Large\bf Acknowledgement}

\ 

I would like to appreciate A. Sugamoto for the 
useful discussions.


\newpage

\begin{thebibliography}{99}
\bibitem{sch}A.S. Schwarz, {\sl Commun. Math. Phys.} 
{\bf 67} (1979) 1; \\
%\bibitem{hs}
G.T. Horowitz, {\sl Commun. Math. Phys.} 
{\bf 125} (1989) 417; \\
%\bibitem{bt}
M. Blau, G. Tompson, {\sl Ann. Phys.} 
{\bf 205} (1991) 130; {\sl Phys. Lett.} {\bf B255} (1991) 
535; \\
%\bibitem{oy1}
I. Oda, S. Yahikozawa, {\sl Phys. Lett.} 
{\bf B234} (1990) 69; {\sl Phys. Lett.} {\bf B238} (1990) 272; 
{\sl Prog. Theor. Phys.} {\bf 83} (1990) 845
\bibitem{pl}J.F. Plebanski, {\sl J. Math. Phys.} {\bf 18} 
(1977) 2511; \\
%\bibitem{ast}
A. Ashtekar, {\sl Phys. Rev. Lett.} {\bf 57} 
(1986) 2244; {\sl Phys. Rev.} {\bf D36} (1987) 1587; \\
%\bibitem{cdj1}
R. Capovilla, T. Jacobson, J. Dell, 
{\sl Phys. Rev. Lett.} {\bf 63} (1989) 2325; \\
%\bibitem{cdjm}
R. Capovilla, J. Dell, T. Jacobson, 
L. Myson, {\sl Class. Quantum Grav.} {\bf 8} (1991) 41; \\
%\bibitem{thft2}
G. 't Hooft, {\sl Nucl. Phys.} {\bf B357} 
(1991) 211; \\
%\bibitem{ike}
H. Ikemori, in {\it Proceedings of the Workshop on 
Quantum Gravity and Topology}, ed. by I. Oda (INS-Report 
INS-T-506)
(1991)
\bibitem{kara}M. Kalb, P. Ramond, {\sl Phys. Rev. } {\bf D9} 
(1974) 2273; \\
%\bibitem{nambu}
Y. Nambu, {\sl Phys. Rep.} {\bf 23C} (1976) 250; \\
%\bibitem{suga}
A. Sugamoto, {\sl Phys. Rev.} {\bf D19} (1979) 
1820; \\
%\bibitem{sos}
K. Seo, M. Okawa, A. Sugamoto, 
{\sl Phys. Rev.} {\bf D19} (1979) 3744; \\
%\bibitem{frtw}
D.Z. Freedman, P.K. Townsend, {\sl Nucl. Phys.} 
{\bf B177} (1981) 282; \\
\bibitem{kkns}M. Katsuki, H. Kubotani, S. Nojiri, 
A. Sugamoto, {\sl Mod. Phys. Lett.} {\bf A10} (1995)2143
\bibitem{jacob} T. Jacobson, {\sl Class. Quantum Grav.} 
{\bf 5} (1988) 923 \\
R. Capovilla, J. Dell, T. Jacobson, 
{\it Class. Quantum Grav.} {\bf 7} (1990) 41
\bibitem{ks} H. Kunitomo, T. Sano, 
{\sl Prog.Theor.Phys.Supplement} {\bf 114} (1993) 31
\bibitem{ezawa} K. Ezawa, Preprint OU-HET/225 (1995) 
hep-th/9511047
%\bibitem{oy2}
%I. Oda, S. Yahikozawa, {\sl Class. Quantum Grav.} 
%{\bf 11} (1994) 2653
%\bibitem{hsin}V. Husain, {\it Phys. Rev. Lett.} {\bf 63} 
%(1994) 800
%\bibitem{},{\it } {\bf } (19)
%\bibitem{},{\it } {\bf } (19)
%\bibitem{},{\it } {\bf } (19)
%\bibitem{},{\it } {\bf } (19)
\end{thebibliography}
\end{document}